\documentstyle{l-aa}      
\begin{document}
\input{psfig.tex}

\thesaurus{03        
           (19.92.1  
            19.94.1  
            07.09.1  
            07.22.1) 
                   }         
\title{The rate of Supernovae from the combined sample of five searches.}

\author{E.\ Cappellaro\inst{1} \and M.\ Turatto\inst{2,1} \and 
D.Yu. Tsvetkov\inst{3} \and O.S. Bartunov \inst{3} \and
C. Pollas\inst{4} \and R. Evans\inst{5} \and M. Hamuy
\inst{6,7}}

\institute{Osservatorio Astronomico di Padova, vicolo dell'Osservatorio 5,
I-35122 Padova, Italy 
\and
European Southern Observatory, Alonso de Cordoba 3107, Vitacura,
Casilla 19001, Santiago 19, Chile
\and
Sternberg Astronomical Institute, Universitetskij Prospect 13,
119899 Moscow, Russia
\and
Observatoire de la C{\^o}te d'Azur, F-06460 Caussols, France 
\and
63 Cassilis Street, Coonabarabran NSW 2357, Australia
\and 
Cerro Tololo Inter-American Observatory, Casilla 603, La Serena, Chile
\and
current address: Steward Observatory, University of Arizona, Tucson,
AZ 85721}

\offprints{E. Cappellaro}

\date{Received ................; accepted ................}

\maketitle

\begin{abstract}
With the purpose to obtain new estimates of the rate of supernovae
we joined the logs of five SN searches, namely the Asiago,
Crimea, Cal{\'a}n-Tololo and OCA photographic surveys and the visual
search by Evans. In this way we improved the SN statistics (the sample
counts 110 SNe) and hence, reduced the uncertainties.

The computation was based on the control time method which allowed the
proper merging of the observations of each galaxy in the various
searches. In addition to discussing the choice for the various input
parameters, we verified the existence of two biases against SN
discoveries, one in the nuclear regions of distant galaxies, most
severe for deep photographic surveys, and a second in inclined spirals,
in particular late spirals.

After correction for these two biases we obtained the rates of each
SN type in the different types of galaxies. We found that the most
prolific galaxies are late spirals in which most SNe are of type II
(0.88 SNu). SN~Ib/c are rarer than SN~Ia (0.16 and 0.24 SNu,
respectively), ruling out previous claims of a very high rate of
SNIb/c. We also found that the rate of SN~Ia in ellipticals (0.13 SNu)
is smaller than in spirals, supporting the hypothesis of different
ages of the progenitor systems in early and late type galaxies.

Finally, we estimated that even assuming that separate classes of
faint SN~Ia and SN~II do exist (SNe~1991bg and 1987A could be the
respective prototypes) the overall SN rate is raised only by 20-30\%,
therefore excluding that faint SNe represent the majority of SN
explosions. Also, the bright SNIIn are intrinsically very rare (2 to
5 \% of all SNII in spirals).

\end{abstract}

\section{Introduction}

The rate of supernovae (SNe) is a key parameter regulating the
chemical evolution of galaxies, the kinematics of the interstellar
medium, the production of cosmic rays, in addition to being a
fundamental constraints for stellar evolution theories.

In recent years there has been a renewed interest in the search for
SNe and in fact the number of discoveries in the last decade almost
equaled those of the previous century.  Despite these efforts,
published estimates of the rate of SN still bring large uncertainties.

In order to obtain a direct measurement of the rate of SNe in a given
galaxy sample one has to divide the number of SN discoveries by the
period of surveillance. The latter can be estimated either by making
some ``reasonable'' assumptions (e.g. Tammann et al. \cite{tamm:94}
and reference therein) or by computing the {\em control time\/}
through the detailed analysis of the log of a given SN search (Zwicky
\cite{zw:38,zw:42}). Since only the SNe discovered in the particular
search enter the computation, in general the main problem with this
approach is the small statistics. This can be improved by joining into
a single database the data of different SN searches but, so far, this
has been done only for the Asiago and Crimea SN searches (Cappellaro
et al. \cite{p1,p2} hereafter PI and PII).

In this paper we present new estimates of the rate of SNe based on the
joint efforts of five SN searches, namely the Asiago
and Crimea searches, the search by Evans (van den Bergh et
al. \cite{vdb:87}, Evans et al. \cite{ev:89}), the OCA search (Pollas
\cite{pol}) and the Cal\'an/Tololo search (Hamuy et
al. \cite{mario:93}). In this way we collected the largest SN sample
ever used for SN rate calculation.

In the following, after introducing the individual searches, we describe
the recipe used in the computation with emphasis on the updating and
improvements with respect to previous works. We make an effort to
explicitly mention all the critical assumptions and parameters
involved in the calculation and to discuss in detail the biases of SN
searches.  Finally, the computed SN rates are reported and compared
with previous estimates.

\section{The SN searches}

In the following we give a brief description of the main
characteristics of each SN search. More details can be found in the
references indicated below.

\begin{description}

\item[Asiago:] the SN search was conducted at Asiago from
 1959 to 1990, initially with the 40/50~cm Schmidt telescope (S40)
and, after 1967, also with the larger 67/92~cm Schmidt (S67). During
the search 31 SNe were discovered (mostly by L. Rosino) and about 20
more, first discovered by others, were recorded in the survey plates
(PI).

\item[Crimea:] this search started in 1961 using the 40~cm astrograph
of the Sternberg Institute in Crimea. It announced the discovery of 21
SNe and 18 more were recorded (Tsvetkov \cite{tsv:83}, PI). The search
is still active but, due to temporary difficulties, few plates
were obtained after 1991.

\item[Evans:] aimed at the prompt discovery of nearby SNe, this  
is the most successful visual SN search. At the beginning, in 1980, a
25~cm telescope was employed which was substituted by a 41~cm
telescope in November 1985 (Evans et al. \cite{ev:89}). Whereas the
search is still active, here we use only the observations performed to
the end of 1988 (almost 100000 individual observations). The SN sample
counts 24 objects (van den Bergh et al. \cite{vdb:87}; Evans et
al. \cite{ev:89}).

\item[OCA:] the search began in 1987 based on the 90/152~cm 
Schmidt telescope of the Observatoire de la C{\^o}te d'Azur (OCA,
Pollas \cite{pol}). Differently from the others, the OCA search is not
systematic but makes use of plates obtained for other purposes.
In particular, there were not predefined sky fields and
observing strategy. Most of the plates were very deep, with limiting
magnitude up to 21-22 and therefore many faint SNe were found. On the
about 500 plates examined to the end of 1994, 68  SNe have been
found (6 were first discovered by others).

\item[Cal\'an/Tololo (C\&T):] this search began in 1990. 
The scientific rationale was to produce a sample of SNe at moderate
distances ($0.01 < z < 0.1$) suitable for cosmological studies.  A
60/90~cm Schmidt telescope was employed for the regular monitoring of
selected fields resulting in the discovery of 49 SNe (Hamuy et
al. \cite{mario:93}) with 5 more  first discovered by others.
\end{description}

With the exception of the visual search by Evans, the searches listed
above use photographic plates and wide field telescopes. The Asiago
and Crimea surveys were aimed at the long term monitoring of
relatively nearby galaxies and were able to discover SNe only up to
the distance of the Coma cluster whereas both the OCA and C\&T
searches could discover SNe up to $z = 0.1-0.15$.

\begin{figure}
\centering
\psfig{figure=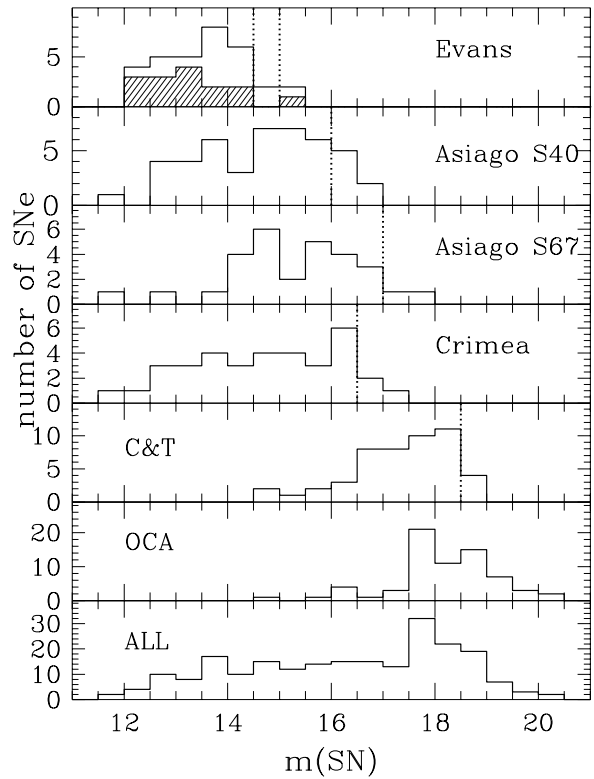,width=9cm}
\caption{Distributions of the apparent magnitudes at discovery,
$m(SN)$, of the SNe found in the different searches. For Asiago,
Crimea and C\&T the magnitudes are B, for
the Evans' search they are visual whereas for the OCA search they are
mixed.  The dashed area in the Evans' search histogram refers to SNe
discovered before 1985 November with the 25~cm telescope. The dotted
lines indicate the adopted limiting magnitudes for each search (for
the OCA search read text).}
\label{limag}
\end{figure}

In Fig.~\ref{limag} we report, for each search, the distributions of
$m(SN)$, the apparent magnitudes of the SNe {\bf at discovery}.  The
cut-off of the distributions at faint magnitudes is imposed by $m_{\rm
lim}$, the limiting magnitudes for SN discovery of the searches which
are, in general, 1-2 mag brighter than the usual definition of plate
limit.  In the bottom panel of Fig.~\ref{limag} is the distribution of
$m(SN)$ for the combined SN sample showing that most SNe have been
found at apparent magnitudes between 13 to 19 with a relatively flat
distribution.

In principle, even for a given telescope and observing strategy
(i.e. plate type, exposure time, etc.) $m_{\rm lim}$ changes from plate
to plate depending on weather conditions but also within the same plate
because of the background variations. For instance, it is most difficult
to discover a SN when it appears projected on the high surface
brightness, inner regions of a galaxy rather than in its outskirts.
However, because of the huge number of observations involved, we
cannot analyze each single case and we are therefore forced to a
statistical approach.  For the systematic surveys
adopting a fixed observing strategy we determined a unique $m_{\rm
lim}$, which is the average of the limiting magnitude for SN discovery
in different plates and plate positions. It must be stressed that even
if the dispersion of the limiting magnitudes of individual observations
can be quite large (of the order of 0.5-1 mag) the assumption of a
unique value for the $m_{\rm lim}$ of a systematic search is not 
critical as long as this value is well determined. A conservative
estimate of the uncertainty of this crucial parameter is $\Delta
m_{\rm lim} = \pm 0.5mag$.

Both in the Asiago and Evans searches two different telescopes have
been used, hence two different $m_{\rm lim}$ have been adopted for
each of the two searches. In particular, for the Evans' search with the
41~cm telescope the value adopted here, $m_{\rm
lim}=15.0$, is 0.4 mag brighter than in van den Bergh \& McClure
(\cite{vdbmc}; see also van den Bergh and Tammann
\cite{vdbt}).

The adopted limiting magnitudes for the systematic searches are
indicated by the dotted vertical lines in the different panels of
Fig.~\ref{limag}.  Note that, allowing for better than average
observing conditions, in each search a few SNe are found at magnitudes
fainter than the average $m_{\rm lim}$.

Because of the lack of a fixed observing strategy, the limiting
magnitude for the OCA observations greatly change from plate to plate.
Fortunately, an estimate of the limiting magnitude for each plate of
this search is available, varying in the range 16 to 21 mag, and has
been used in the calculation.

\section{Computational recipe}

The present determination of the rate of SNe is based on the {\em
control time} method which has been introduced by Zwicky (\cite{zw:42}) and
subsequently revisited by several authors (e.g. Cappellaro \& Turatto
1988, PI, PII, van den Bergh et al. \cite{vdb:87}, van den Bergh \&
McClure \cite{vdbmc}). In the following we discuss the assumptions
and the algorithm adopted in the calculation.

\subsection{Galaxy sample}

With the exception of the Evans' search, the SN searches that we are
considering are based on wide field plates which, in a single shoot,
allow the surveillance of many galaxies. Therefore
the galaxy sample is not defined {\em a priori} but was selected
by extracting from a suitable list those galaxies
which appear in at least one of the survey plates.  As we will
describe later on, for the  computation we need
to know for each galaxy the recession velocity, the morphological
type, the luminosity and, for spirals, the axial ratio.  Hence we need
a list of galaxies for which these data are available
and homogeneous.

In analogy to PII we use as input list the Third Reference Catalogue
of Bright Galaxies (de Vaucouleurs et al. \cite{rc3}, hereafter RC3).
It turns out that, among the about 23000 galaxies listed in the RC3,
over 10000 have been observed at least once by our surveys.
Unfortunately, for many of them some of the required parameters are
missing in the RC3.  To reduce this problem, we consulted the Leda
extragalactic database\footnote{The Lyon-Meudon Extragalactic Database
(LEDA) is supplied by the LEDA team at the CRAL-Observatoire de Lyon
(France).} and integrated missing data when available. After that,
galaxies still with incomplete data were excluded from the sample
(7773 galaxies remaining).

Obviously in the computation of the SN rate only the SNe discovered in
one of the galaxies of the sample have to be considered.  This reduces
the number of SNe of our sample from 211, the total number of SNe
discovered in the five searches, to 110.  The reduction is especially
severe for the OCA and C\&T searches whose targets are distant
galaxies most of which are not listed in the RC3 or for which the
required data are not available, and instead has no effect for the
Evans' search.

\subsection{SN light curves}\label{mt}

For each galaxy of the sample, we need to compute $m_{\rm
sn}(t)$, the ``apparent'' light curve of a possible SN in that
galaxy. This depends on the SN type and on the galaxy distance
or, more precisely, we can write:

$$ m_{\rm sn}(t) = M_{\rm 0,sn}+\Delta m_{\rm sn}(t)
+\mu+A_g+<A_i> $$

where $M_{\rm 0,sn}$ is the intrinsic (extinction corrected) absolute
magnitude at maximum of the SN, $\Delta m_{\rm sn}(t)$ describes the
light curve evolution relative to maximum, $\mu$ is the galaxy
distance modulus and $A_g$ the galactic absorption. Finally, $<A_i>$
is the average extinction of the SN population in that galaxy due to
the internal dust as seen from our particular line of sight.

Similar to PI and PII, $\mu$ was derived from the Hubble flow
velocity ($v3k$ in RC3) if this was larger than 1500 km~s$^{-1}$,
otherwise the distance modulus from Tully's catalog (Tully
\cite{tully}) was adopted\footnote{Throughout this paper $H_0=75\,{\rm
km}\,{\rm s}^{-1}\,{\rm Mpc}^{-1}$ is assumed.}. For the galactic
absorption, $A_g$, we used the Burstein \& Heiles estimates as
reported in the RC3.

The light curve shapes, $\Delta m_{\rm sn}(t)$, of the different types
of SNe were obtained by selecting templates from literature
(Leibundgut et al. \cite{bruno}, Patat et
al. \cite{patat1,patat2}). Because of the different kinds of searches
involved we needed templates for B, V and R bands.

In the present work we calculate the rates for the three basic types
of SNe namely Ia, Ib/c and II.  At present the statistics are not large
enough to separate Ib from Ic that, for this reason, were lumped
together. In PI the adopted templates for SN~Ia and SNIb/c were the
same (although the absolute magnitudes at maximum were
different). Instead in the present work we adopt the light curves of
SN~1990I (Della Valle et al., in preparation) as template for SN~Ib/c.
Concerning SN~II, it is well known that they exhibit quite
heterogeneous photometric behaviors with different light curve shapes
and maximum luminosities (cf. Patat et al. \cite{patat1,patat2}). To
account for this we calculate separately the rates for the two
photometric classes of IIP (plateau) and IIL (linear) and derive the
total rate of SNII by summing the two contributions. It turns out that
the results are very similar to those which are obtained by adopting,
in the calculation of the control time, a light curve which is
intermediate between Plateau and Linear. No account is made for the
rare class of SN~IIn or for the possible existence of a separate class
of faint SNII similar to SN~1987A. Some comments about the relative
contribution of these classes of SNe will be made in the discussion.

More tricky is the problem of determining $M_{\rm 0,sn}$ and $<A_i>$,
the latter being, in general, unknown. Let us discuss this 
important point in some detail.

In a first approximation, we can assume that ellipticals are dust
free, that is $<A_i>=0$ mag. Therefore by taking the average absolute
magnitude of SN~Ia in ellipticals (Ia are the only type of SNe found
in these galaxies) we obtain directly $M_{\rm 0,Ia}$. In particular,
we adopt the value reported by Miller \& Branch (\cite{miller}), $<
M_{\rm B}> = -18.95$ mag.

Dust extinction is certainly important in spiral galaxies.
For the moment let us neglect the dependence of $A_i$ on the
inclination of the disk along the line of sight, which will be
discussed in Sec.~\ref{incsec}.  Direct estimates of the average
absorption suffered by SN~Ia in face-on spirals range from
$<A^B_{i,0}> = 0.4$ mag (Miller \& Branch \cite{miller}) to $ 0.7$ mag
(Della Valle and Panagia, \cite{mdv:92}), that is the average observed
magnitude of SN~Ia is 0.4-0.7 fainter than the intrinsic value.

On the other side there are indications that the intrinsic magnitude
of SN~Ia correlates with the Hubble type of the parent
galaxies. Evidence are still preliminary but it seems that in spirals
$M_{\rm 0,Ia}$ is 0.3--0.5 mag brighter than in ellipticals (Branch et al.
\cite{branch}, Hamuy et al. \cite{mario:95}).  Given the uncertainties, we will
assume that the two effects cancel out and that the ``observed''
absolute magnitudes ($<M_B>=M_{\rm 0}+<A_i>$) of the SN~Ia in
ellipticals and face-on spirals are the same.

For type II (either P or L) and Ib/c there are no estimates of the
average absorption in the parent galaxies. We simply assume $<M_B>$
from Miller \& Branch (\cite{miller}) as representative of the ``observed''
absolute magnitudes for these classes of SNe in face-on spirals,
$<M_{B}> = -17.11$, $-16.53$ and $-17.05$ for Ib/c, IIP and IIL,
respectively. The above assumption is probably reasonable given that,
although the Miller \& Branch sample includes SNe in spirals of all
inclinations it certainly suffers a  bias against heavily
absorbed SNe.

Similar to PI we allowed for a Gaussian distribution of the SN
magnitudes with variance $\sigma$ corresponding to the observed
ones.  We stress that, as demonstrated in PI, the calculated SN rate
is not very sensitive to changes in the shapes or width of the SN
luminosity distribution, but instead it strongly depends on the mean
value.

\begin{figure}
\centering
\psfig{figure=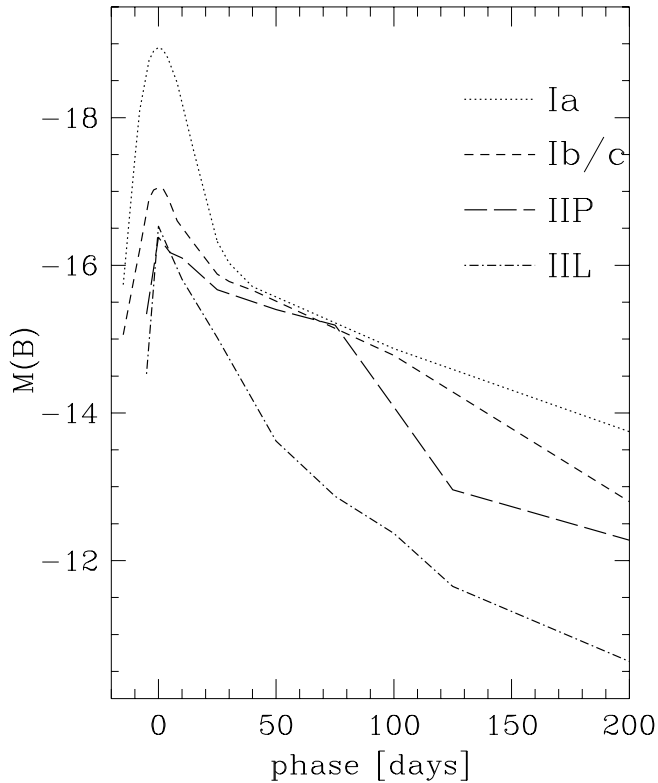,width=9cm}
\caption{B template light curves.
For the calculation of the control time, the light curves are
truncated 400 days after maximum.}
\label{lc}
\end{figure}

The need for an analytical approximation of SN light curves (cf.
PI) has been eliminated and observed light curves can be readily
input to the program.  The adopted templates for the B band are shown
in Fig.~\ref{lc}. 

\subsection{Control time and SN rate}

Once we compute for each galaxy of the sample the expected
light curves, $m_{sn}(t)$, of all SN types, we can obtain the control
time $ct_{\rm i,j}^{\rm sn}$, that is the interval of time during
which a possible SN in the $j^{th}$ galaxy stays brighter the $m_{\rm
lim}$ of the $i^{th}$ observation.  Next we compute $tct_{\rm j}^{\rm
sn}$, the total control time for the series of $n$ observations of the
galaxy (sorted on the epoch $t_i$) according to the following recipe:

\begin{equation}\label{tct}
 tct_{\rm j}^{\rm sn} = \sum_{\rm i=1}^{\rm n}{\Delta t_{\rm i} \times
L_j c_i} 
\end{equation}

where:

\[ \Delta t_{\rm i} = \left\{ \begin{array}{lc} 
     ct_{\rm i,j}^{\rm sn}~~~~~~~ &  \mbox{if~~ $t_{\rm i}-t_{\rm i-1}\ge 
ct_{\rm i,j}^{\rm sn}$ ~~or~~~ $i=1$} \\ 
                 t_{\rm i}-t_{\rm i-1} & \mbox{otherwise}~~~, 
                              \end{array} 
                   \right. \]

$L_j$ is the galaxy blue luminosity and $c_i$ is a correcting factor
introduced to account for the bias against SN discovery in the nuclear
regions of galaxies which will be discussed in Sect.\ref{secshaw}. The
galaxy luminosity is introduced as a normalization factor because it has
been demonstrated that the SN rate is proportional to the galaxy
luminosity (Tammann \cite{tamm:74}; PII). $L_j$ is computed from the
$B_{\rm T}^0$ magnitude reported in the RC3 and expressed in unit of
solar luminosity adopting $M^B_\odot = 5.48$.  

Finally, we compute $\nu_{\rm sn}$, the rate of a given SN type in each
galaxy sample as:

$$ \nu_{\rm sn} = \frac{N_{\rm sn}}{\sum_{\rm j=1}^{\rm N_G} tct^{\rm
sn}_{\rm j}} $$

where $ N_{\rm sn} $ is the number of SNe of that type discovered
during the search(es) in the $N_G$ galaxies of the sample. The rate
$\nu_{sn}$ is expressed in SNu, i.e. SNe per $10^{10}$ L$_\odot$ per
century and, because of the normalization to galaxy luminosity, scales
as $(H/75)^2$.

For about 1/4 of the SNe of our sample no detailed classifications are
available.  Most of them are of type I but for a few (10 out of 110)
not even this broad classification is available. The unclassified SNe
have been redistributed among the three basic types according to the observed
distribution  in the merged sample, that is
in E-S0 100\% type Ia, in spirals 35\% type Ia, 15\% type Ib, 50\%
type II.

\section{The SN rates from individual SN searches}

It is a useful exercise to compare the rate of SNe calculated using
separately the data of each search.  The results are shown in
Tab.~\ref{indiv} where for each SN search we report the number of RC3
galaxies included in the search (col.~2), the average redshift of the
galaxies of the sample (col.~3), the total control time for SN~Ia
(which gives an indication of the weight of the search on the combined
sample) (col.~4), the number of SNe discovered in the RC3 galaxy
sample (col.~5), and the overall SN rate (col.~6) before corrections
for the biases which will be discussed in
Sect.~\ref{sele}. Note that the numbers in the last row giving the
results for the combination of the five searches are not the plain
sums of the values in the columns since several SNe and galaxies are in
common between individual searches and because, according to the
definition, the cumulative control time is smaller (or equal if the
searches do not overlap) than the sum of the control times of the
individual searches.

\begin{table*}
\caption{The average rate of SNe [SNu] for each SN search.}\label{indiv}
\begin{tabular}{lcccccll}
\hline
search & number of & $<v>$ & TC(Ia)~~~ & number &
\multicolumn{3}{c}{SN rate [SNu]} \\ 
\cline{6-8}
       & galaxy    & km s$^{-1}$ & $100 yr \times {10^{10}\rm L}_{\sun} $ & of SNe &
~~raw~~ & $-$ nuclear bias  & $-$ inclination bias\\
\hline
Evans  &1377 &1805 &72~~~ & 24~~~& 0.54 & 0.54               & 0.80 (0.69 - 0.88)\\
Asiago &2412 &3960 &197~~& 51~~~& 0.39 & 0.50 (0.42 - 0.58) & 0.76 (0.65 - 0.86)\\
Crimea &2697 &3603 &112~~& 33~~~& 0.41 & 0.50 (0.45 - 0.53) & 0.76 (0.65 - 0.86)\\
OCA    &4352 &6200 &72~~~ & 16~~~& 0.29 & 0.46 (0.35 - 0.63) & 0.58 (0.54 - 0.61)\\
C\&T   &1775 &5040 &60~~~ & 12~~~& 0.21 & 0.34 (0.26 - 0.44) & 0.44 (0.40 - 0.47)\\
\\
All    &7773 &4320 &425~~&110~~& 0.35 & 0.45 (0.39 - 0.52) & 0.66 (0.58 - 0.74)\\
\hline
\end{tabular}
\end{table*}

The raw rates of the different searches reported in col. 6 differs by
more than a factor 2.  In the next section we will show that
observational biases affect more heavily photographic surveys and,
among these, those aimed to reach fainter limiting
magnitudes. Therefore, the fact that the Evans' visual search gives
the highest raw SN rate whereas the deep OCA and C\&T searches the
lowest values, is not unexpected.

\section{Biases on SN discovery}\label{sele}

\subsection{The nuclear regions of galaxies}\label{secshaw}

It was first pointed out by Shaw (\cite{shaw}) that in the general
list of SNe there is an apparent deficiency of objects in the inner
regions of more distant galaxies compared with nearby ones. To some
extent this is due to a bias in the parent galaxy sample, more distant
SN parent galaxies being in the average bigger (Cappellaro \& Turatto,
\cite{ct:96}), but even after this dependence has been removed by
normalizing the SN radial distances to the galaxy radius, it results
that at least 40\% of the SNe which explode in galaxies with recession
velocities larger than 3000 $\mbox{km s$^{-1}$}$ are lost because of
this effect.  The bias is found to be more important for deep
photographic searches and negligible for visual and CCD searches in
nearby galaxies.

\begin{figure}
\centering
\psfig{figure=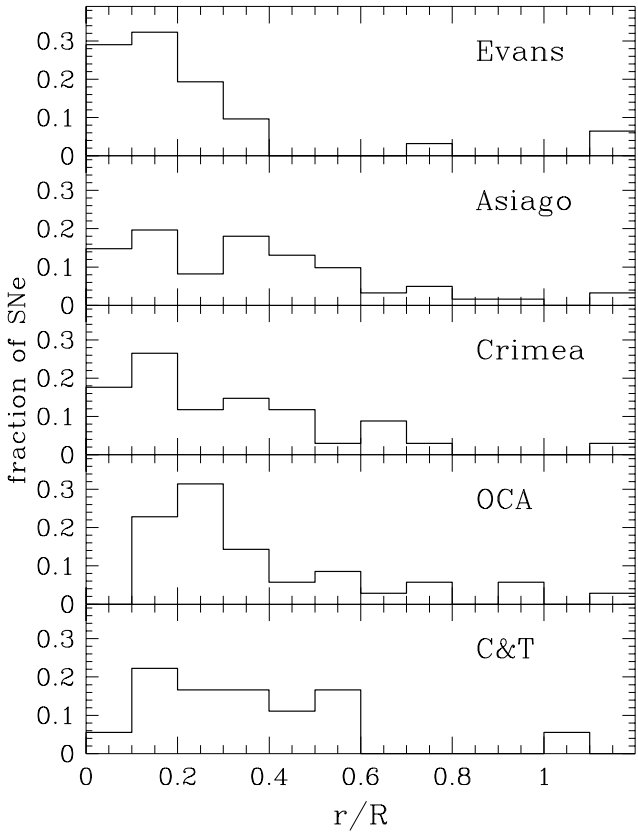,width=9cm}
\caption{In the different panels we report for each search the
distributions of $r/R$, the ratio of the distances
from the nucleus of the SNe discovered to the semi-major axis
of the parent galaxies. To improve the statistics, for the Evans sample
we included SNe discovered after 1989.}\label{radr}
\end{figure}

To verify this effect for the search we are considering, in
Fig.~\ref{radr} we compare the relative radial distributions of the
SNe found in the different searches. As can be seen the peak of
the observed distribution shifts from the galaxy nucleus to an outer
radius moving from the Evans' visual search to the photographic Asiago
and Crimea surveys and even more outward for the deep OCA and C\&T
surveys.

In some cases, especially for very deep exposures, the nuclear bias
can be due to the over-exposure on the photographic plates of the high
surface brightness, inner regions of the galaxies. Even if the
saturation regime is not reached, as was normally the case for the
searches we considered, it is most difficult to identify a new object if
it appears projected on regions of high photographic density. Also,
the small scale of the telescopes usually employed for photographic
surveys hampers the discovery of a SN when it projects near the galaxy
nucleus.  Visual and CCD searches are less biased because of a better
dynamic range and, typically, larger scale.

For our purposes the consequence of this bias is that a portion of
each galaxy cannot be probed by photographic SN searches. We accounted
for it by introducing in equation~(\ref{tct}) an appropriate correcting
factor, $c_i$, measuring the fraction of the galaxy luminosity,
$L_j$, actually surveyed by the $i^{th}$ observation. 

The value of the correcting factor, $c_i$, and its dependence on
the search and galaxy type and distance were estimated in two independent
ways:

\begin{enumerate}
\item
we selected from the RC3 sample a few galaxies of different types,
luminosities and recession velocities for which detailed surface
photometry was available (e.g. Caon et al. \cite{caon}, J{\o}rgensen
et al. \cite{jorg}). Then we retrieved from the archives of the
Asiago, Crimea and OCA searches a few typical survey plates on which
these galaxies appear. Via artificial star experiments (cf. Turatto et
al. \cite{tur:94}) and direct examination of the plates we estimated
for each galaxy the limiting radial distance within which SNe
cannot be detected. With this, and knowing the luminosity profiles, we
estimated the fraction of the galaxy which is unaccessible to the
search. The results of this analysis can be summarized as follows: i)
whereas we found significant differences from case to case, there is
no clear evidence of an increase of the bias with the galaxy distance;
ii) in the case of early type galaxies the fraction of the galaxy
unaccessible to the search is in the average 10-20\% for the Asiago
and Crimea searches, 20-40 \% for the OCA search; iii) the bias seems
less severe in spirals than in ellipticals (roughly by a factor 2) .

\item we can estimate the importance of the nuclear bias also by
assuming that the bias is negligible for the visual search in nearby
galaxies. In particular we assume that the radial distribution of the
Evans' sample gives the intrinsic radial distribution of the SNe.
Instead the radial distribution of the SNe found in photographic
searches can be considered unbiased only outside a given radius which
we adopted $r/R=0.3$. Thus the fraction of SNe which is lost in
photographic surveys can be estimated by normalizing the radial
distributions of the different searches for $r/R>0.3$.

With this method we obtain correcting factors which are significantly
larger that those obtained with the former one. Among the photographic
surveys, the least affected appears the Crimea search (with an
estimated SN deficiency of $\sim 25\%$), more biased the Asiago search
($\sim 35\%$) and the OCA and C\&T searches ($\sim 50\%$). There is
no apparent difference between early and late type galaxies but the
SN statistics for early type galaxies are very poor. It appears that there is
a small trend with the distance, the bias in galaxies with recession
velocities greater than $3000$ km~s$^{-1}$ being $20-30\%$ more severe
than in nearby ones.
\end{enumerate}

Since we have no reasons to favor either result, we decided to adopt
as correcting factors the averages of those estimated in the
two ways described above. The values of $c_i$ range from 0.85 for the
Crimea observations of nearby spiral galaxies to 0.55 for the OCA and
CTIO observations of distant ellipticals. No corrections was introduced
for the Evans observations ($c_i = 1$).

The values of the total SN rates for each SN search corrected for the
nuclear bias are reported in col.~7 of Tab.~\ref{indiv} where
in parenthesis are reported the extreme values obtained using
alternatively the correction factors of either method. These give an
indication of the uncertainties of the correction.

As expected, after this correction the average SN rates of
the different searches become very similar with the exception of the
C\&T  rate for which the small statistics may be a problem.

\subsection{The inclination of spirals}\label{incsec}

The presence of a strong bias in the discovery of SNe in inclined
spiral galaxies has been first mentioned by Tammann
(\cite{tamm:74}). Subsequently several authors (cf. van den Bergh \&
Tammann \cite{vdbt}, PII) have tried to give an estimate of the
importance of this bias and of the possible dependence on the search
and galaxy parameters.

In particular it has been claimed that the effect is negligible for
CCD and visual searches though, because of the small statistics, this
was not a firm conclusion.  To test this point we divided our combined
galaxy sample in two bins of low and high inclinations\footnote{The
inclination ($\alpha$) of spirals is derived from the axial ratio
$R_{25}$ as reported in RC3 catalog using the relation $\alpha =
\arccos 1/R_{25}$.} and calculated the SN rates separately for the
Evans' visual search and for the combined photographic searches. The
results are shown in the first two rows of Tab~\ref{inctab}, in which
in columns 2 and 3 are the SN rates in galaxies of different
inclination and in col.~4 their ratio.  Since the statistical errors
are large no definite conclusion can be drawn, but the indication is
that the bias affects also the visual search although to a lesser
extent than photographic searches.

Based on the general list of SNe, van den Bergh (\cite{vdb:91}) argued
that the bias due to inclination is most severe for SNII and
negligible for SNIa.  In PII we found that the bias was roughly a
factor 2 larger for type II than for type Ia. Actually, on our present
larger sample we find no significant difference between SN~Ia and
SNII+Ib (Tab.~\ref{inctab}).

We investigated also the dependence on the galaxy morphological types
and found that the bias is stronger in late $Sb-Sd$
galaxies than in early $S0-Sa$ spirals (Tab.~\ref{inctab}).

\begin{table}

\caption{Comparison of the SN rates in spiral galaxies with small
($\alpha<45^\circ $) and 
large ($\alpha \ge 45^\circ $) inclination.}\label{inctab}

\begin{tabular}{lrrr}
\hline
                & \multicolumn{2}{c}{SN rate [SNu]}\\
                & $\alpha<45^\circ$   & $\alpha\ge 45^\circ$ & ratio~~\\
\hline
Evans' visual   & $0.85\pm0.27$ & $0.56\pm0.16$  & $1.5\pm0.6$\\
photographic    & $0.76\pm0.12$ & $0.39\pm0.06$  & $1.9\pm0.4$\\
\\
Ia              & $0.20\pm0.04$ & $0.11\pm0.02$ & $1.8\pm0.5$\\ 
II+Ib           & $0.53\pm0.11$ & $0.30\pm0.06$ & $1.8\pm0.5$\\ 
\\
S0-Sab          & $0.34\pm0.10$ & $0.28\pm0.07$ & $1.2\pm0.5$\\
Sb-Sd           & $0.96\pm0.16$ & $0.47\pm0.08$ & $1.9\pm0.5$\\
\\
All - no corr.  & $0.73\pm0.11$ & $0.41\pm0.06$ & $1.8\pm0.4$\\
All - $\sec i$  & $0.76\pm0.12$ & $0.66\pm0.09$ & $1.2\pm0.2$\\
All - adopted   &$0.87\pm0.13$& $0.83\pm0.11$ & $1.0\pm0.2$\\
\hline
\end{tabular}
\end{table}

In previous works (Cappellaro \& Turatto, \cite{ct88}; PII) the inclination
bias was corrected adopting a purely empirical approach, that is
simply multiplying the SN rate in inclined spirals by a proper factor
so that the dependence of SN rate from galaxy inclination disappears.

Here we have adopted a different approach, assuming that the reduced
efficiency of SN searches in inclined spirals is due to the large
absorption suffered by SNe in these galaxies.  In passing we note that
since extinction is smaller for longer wavelengths this naturally
implies a less severe bias for observations in the visual and red
bands than for observations in the B band.

In Sect.~\ref{mt} we already accounted for the internal absorption
suffered by SNe in face--on galaxies. In addition, if both dust and SNe
are distributed in slabs with common median plane, the average SN in a
galaxy of inclination $\alpha$ will experience the additional
absorption $A_{i,0}\times (\sec\alpha -1)$, where $A_{i,0} = 1.086
\times \tau_0$ and $\tau_0$ is the half-width optical depth of the
dust layer.  The value of $\tau_0$ is still debated but typical values
are $2\tau_{0,B} \sim 1$ (Bottinelli et al. \cite{botti}). A similar
value ($2\tau_0^B = 1.28$) was reported by Della Valle \& Panagia
(\cite{mdv:92}) based on the direct measurements of the average
extinction suffered by SNIa.

These estimates are appropriate for $Sb-Sc$ galaxies whereas
extinction is probably smaller in early $S0-Sa$ spirals (Valentijn
\cite{valen}). Based on these considerations we adopt $A_{i,0}^B =
0.7$ for $Sb-Sd$ galaxies and $A_{i,0}^B = 0.35$ for the other
spirals.

\begin{figure}
\centering
\psfig{figure=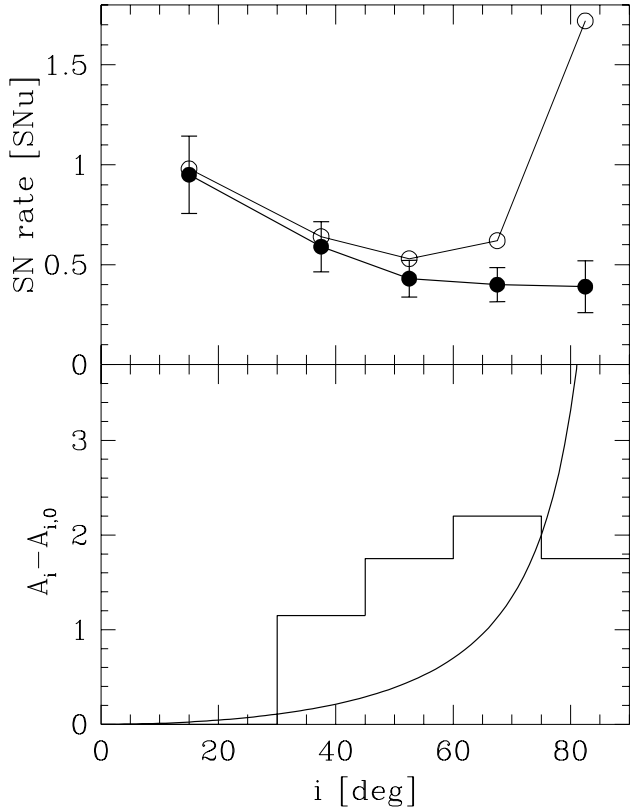,width=9cm}
\caption{Upper panel: the SN rates in spirals of different
inclination as observed (filled circles) and after the $A_{i,0}
(\sec\alpha-1)$ correction has been applied (open circles).  Error-bars
relative to the SN statistics are indicated.  Bottom panel: comparison
between the $A_{i,0} (\sec\alpha-1)$ extinction law and the extinction
law which eliminates the dependence of the SN
rates on galaxy inclination. }
\label{inc}
\end{figure} 

Once the correction for the inclination effect is included in the
computation, the ratio of the SN rates in galaxies with inclination
$\alpha<45^\circ$ to that of more inclined galaxies decreases from
$1.8\pm0.4$ to $1.2\pm0.3$, consistent with the expected intrinsic
value of 1 (Tab.~\ref{inctab}).  However a closer examination reveals
that the correction is too small for galaxies of intermediate
inclination and too large for edge-on spirals. This is
better seen in the upper panel of Fig.~\ref{inc} where we compare the
SN rates in spiral galaxies of different inclinations before and
after the afore mentioned correction.  Changing the value of
$A_{i,0}$ would not help in this respect, since the problem is the
functional dependence on $\sec\alpha$.

Thus we reversed the problem asking which functional dependence of
$A_i$ on inclination would give the appropriate correction. In
practice we have determined for each inclination bin the value of the
absorption needed to cancel the dependence of the SN rate on
inclination.  The result of this test is shown in Fig.~\ref{inc} and
can be interpreted in two ways, either the inclination bias is due not
only to extinction in the parent galaxies but also to some additional
effect (e.g. related to the galaxy surface brightness) or the
assumption of a plane parallel geometry for the dust distribution has
to be rejected.

The latter interpretation was favored also by van den Bergh
(\cite{vdb:91}) based on the observed distribution of SN in inclined
galaxies. He suggested that SNe, in particular SN~II in $Sc-Sd$
galaxies, explode at the bottom of chimney-like dust structures.  In
this scenario the value of $A_i$ as a function of inclination depends
on the details of the geometry but, schematically, we expect the
extinction to be negligible for small inclination and suddenly to rise
when the line of sight to the SNe intersects the wall of the chimney.
Actually the empirical extinction law shown in the bottom panel of
Fig.~\ref{inc} could better be explained by a model in which SNe are
associated to dust clouds which are elongated along the disk of the
galaxy.  The fact that the extinction peaks for galaxy inclination
around $60^\circ$ could suggest an axial ratio for the clouds of the
order of two.  This scenario is consistent with other recent
findings. From one side Beckman et al. (\cite{beck}) showed that dust
in spirals has an irregular distribution with an optical depth one
order of magnitude larger in the spirals arms than in the inter-arm
regions. On the other side SNII, Ib/c and at least half of SN~Ia are
found to be associated with the spiral arms (Della Valle \& Livio
\cite{mdv:94}, Bartunov et al.  \cite{bartunov}), therefore they occur
in regions of higher than average extinction.

Conservatively, we will adopt as the functional dependence of the
extinction on galaxy inclination the average of the $\sec\alpha$ and
of the ``empirical'' extinction laws reported in Fig.~\ref{inc}.  The
SN rates for each SN search after this correction are reported in the
last column of Tab.~\ref{indiv}. Again the values obtained using
alternatively one of the two extinction laws are reported in
parenthesis to give an indication of the errors.

We stress that the estimate of the SN rate in spirals, in particular
late spirals, are quite sensitive to the adopted value of the average
extinction in galaxies. For istance if we adopt for $Sb-Sd$ $A_{i,0}^B
= 0.35$ (instead of 0.7) the estimate of the SN rate in these galaxies
decreases by about 40\%. With this choice however the
inclination bias in not corrected, that is the rate of SNe in
spirals with $\alpha<45^\circ$ remains 1.5 larger than in
more inclined spirals.

\section{SN rates and uncertainties}

\begin{table*}
\caption{The rate of SNe in SNu corrected for selection
effects.}\label{final}
\begin{tabular}{lcccccccc}
\hline
galaxy &
\multicolumn{3}{c}{No. of SNe} &&
\multicolumn{4}{c}{SN rate [SNu]} \\
\cline{2-4}\cline{6-9}
type    &  Ia  & Ib   &  II   &&  Ia  &  Ib      & II      & All  \\
\hline
E       &  7.0 &      &       && 0.13  &$\le0.03$&$\le0.04$& 0.13 \\    
S0      &  9.0 &      &       && 0.18  &$\le0.04$&$\le0.04$& 0.18 \\
S0a-Sa  &  7.1 & 2.4  &  1.5  && 0.25  & 0.16    & 0.16    & 0.57 \\
Sab-Sb  &  8.7 & 2.3  & 11.0  && 0.17  & 0.09    & 0.53    & 0.78 \\
Sbc-Sc  & 14.6 & 5.9  & 16.5  && 0.23  & 0.20    & 0.75    & 1.19 \\
Scd-Sd  &  5.0 & 0.5  & 10.5  && 0.24  & 0.04    & 1.22    & 1.49 \\
Others  &  3.7 & 1.8  &  2.5  && 0.27  & 0.22    & 0.41    & 0.90 \\ 
\\
E-S0    & 16.0 & 0.0  &  0.0  && 0.15  &$\le0.02$&$\le0.02$& 0.15 \\
S0a-Sb  & 15.8 & 4.7  & 12.5  && 0.20  & 0.11    & 0.40    & 0.71 \\
Sbc-Sd  & 19.7 & 6.3  & 27.0  && 0.24  & 0.16    & 0.88    & 1.27 \\
\hline
\end{tabular}
\end{table*}

\begin{figure}
\centering
\psfig{figure=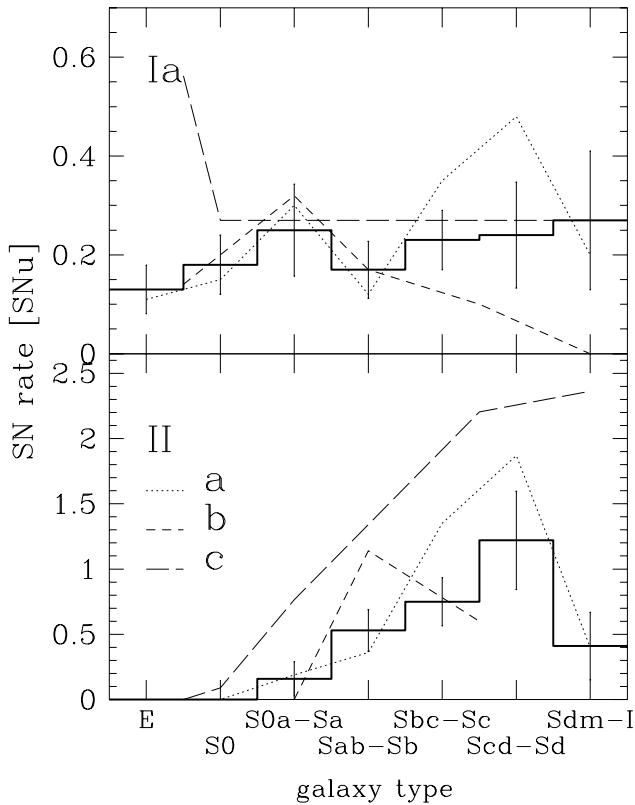,width=9cm}
\caption{The rate of SNIa and SNII in the different type of galaxies
(upper and lower panel, respectively). SNIb/c are not reported because
of the poor statistics (see Tab.~3). Error-bars due only to SN
statistics are indicated. For comparison we report the rates of PII
(a), van den Bergh \& McClure 1994 (b) and Tammann et al. (1994) (c),
scaled to the value of the Hubble constant adopted here (H=75 km
s$^{-1}$ Mpc$^{-1}$).}
\label{finfig}
\end{figure}

With the recipe, the parameters and the correcting factors discussed
above we were finally able to compute the SN rates for the combined
sample of the five SN searches. The results are reported in
Tab.~\ref{final} and the dependence of the SN rates on galaxy morphology
is displayed in Fig.~\ref{finfig} for SN~Ia and SN~II (the statistics
for SN~Ib are too small). For comparison, recently published estimates
are also reported. In general the present results are consistent with
the previous estimates, with a few distinctions.

We find that the rate of SNIa increases by about a
factor 2 moving from ellipticals to late spirals. This confirms
the results of PII but is in contradiction with those of
Tammann et al. (\cite{tamm:94}).
Actually, our results are in excellent agreement with those of
Tammann et al. for the SNIa rate in spirals (note however
their {\em a priori} assumption of a constant rate from early
to late spirals) but our estimate of the rate in ellipticals is a factor 
4 smaller. As discussed in Turatto et al. (\cite{tur:94}) we
believe that the disagreement is due to a bias of SN discoveries in
the ``fiducial sample'' of galaxies.

Traditionally, the discovery of SNIa in ellipticals has been the main
argument for placing the progenitors of SNIa among old stellar
population.  On the other side the fact that the rate in late spirals,
which are dominated by stars of population I, is twice that in ellipticals
implies that the average age of the progenitors of SNIa in spirals is
younger.

It might be expected that the different ages of the progenitor systems
lead to observable differences in the outcomes of the explosions.
Indeed there are increasing evidences that the variance of SNIa
properties in spirals is significantly larger than in ellipticals
(Filippenko \cite{fil:96}).  Therefore our results support the
idea that the progenitors of SN~Ia in spirals can have
quite different ages.

We also confirm (cf. PII) that SNIb in late spirals are ``only'' 40\%
of all type I SNe, ruling out definitively the very high rate
suggested by Muller et al. (\cite{muller}).

The rate of SNe, in particular SNII, is found to peak in the late
$Scd-Sd$ spirals instead that in $Sab-Sb$ as indicated by van den
Bergh \& McClure (\cite{vdbmc}). The odd result by van den Bergh \&
McClure is most likely due to the small statistics but differences in
the galaxy catalog might also contribute.  In fact their reference
catalog was the Revised Shapley-Ames Catalogue (Sandage \& Tammann,
\cite{rsa}) while we used the RC3.

Finally we point out that the present estimate of the overall SN rate
in late spirals is $\sim 40$\% lower than in PII, although the
difference is within the errors. This is mainly due to the smaller
correction factors adopted here for the galaxy inclination bias. We
should stress that because several galaxies have independent
observations in different searches, the dependence of the new SN rates
on the adopted input parameters and bias corrections is reduced
compared with PII.

In most of the previous works, SN rates are presented only with errors
due to SN statistics since, because of the poor statistics, these
errors dominated over others. In our case, the enlargement of the SN
sample strongly reduces the relative importance of statistical errors
and other contributions become important, in particular when
considering broad bins of galaxy morphological types.  In
Tab.~\ref{error} we have summarized the different components and the
total errors calculated as follows:

\begin{enumerate}
\item statistical errors are those due to the statistics of the SN
events which are assumed to conform to a Poisson distribution.

\item parameter errors are estimated from the differences of the
SN rates computed using extreme values of the input parameters. The
cumulative errors due to the parameters are calculated by adding
quadratically the individual contributions, assuming that they are
uncorrelated. We included an uncertainty of $\pm0.25$ mag on the SN
absolute magnitudes, $\pm 0.5$ mag on the adopted limiting magnitude
of each search, and $\pm 50\%$ errors on the absolute magnitude
dispersions. For the uncertainties on the shapes of the light curves
we adopted for type I the difference between Ia and Ib light curves
and for type II the difference between IIP and IIL.  We
neglected the errors on the galaxy data, that is on radial velocities,
luminosities, etc.

\item both the corrections for the nuclear and for the
inclination biases have been determined using the average between the
correcting factors estimated with two different methods
(cf. Sec.~\ref{sele}).  As a measure of the propagated uncertainties
of each correction we have adopted the difference between the SN rates
calculated applying either method.
\end{enumerate}

Finally the total errors are estimated by adding quadratically
the three components.  As can be seen from Tab.~\ref{error},
once the galaxies have been grouped in broad morphological bins
the contributions of the three sources of errors are
similar.  By comparing Tab.~\ref{error} with the similar table of PII
it appears that the relative errors of the present estimates are
reduced, in the average, by about 20\%.

\begin{table*}
\caption{The errors of the rate of SNe (in SNu)}\label{error}
\begin{tabular}{l|ccc|ccc|ccc|ccc}
\hline
galaxy &
\multicolumn{3}{c|}{statistical}&
\multicolumn{3}{c|}{parameters}&
\multicolumn{3}{c|}{biases }&
 \multicolumn{3}{c}{SN rate [SNu]}\\
type    &  Ia  &  Ib/c  & II 
        &  Ia  &  Ib/c  & II    
        &  Ia  &  Ib/c  & II    
        &  Ia  &  Ib/c  & II    \\
\hline
E-S0    & 0.04 &      &      & 0.03 &      &      & 0.02 &      &
        &$0.15\pm0.06$&      &       \\
S0a-Sb  & 0.05 & 0.05 & 0.11 & 0.04 & 0.03 & 0.12 & 0.04 & 0.02 & 0.09
        &$0.20\pm0.07$& $0.11\pm0.06 $ & $0.40\pm0.19$\\
Sbc-Sd  & 0.05 & 0.06 & 0.17 & 0.05 & 0.04 & 0.24 & 0.05 & 0.04 & 0.22 
        &$0.24\pm0.09$& $0.16\pm0.08 $ & $0.88\pm0.37$\\
\hline
\end{tabular}
\end{table*}

Using the SN rates given in Tab.~\ref{final} we can compute the
expected rate of SNe for any given galaxy (or galaxy sample) for which
morphological type and luminosity are known. In particular we can estimate
the expected rate of SNe in the Galaxy with the assumption that the
morphological type is $Sb$ and the luminosity is $2\times10^{10}$
L$_\odot$. In a millennium we expect $4\pm1$ SN~Ia,
$2\pm1$ SN~Ib/c and $12\pm6$ SN~II, where the uncertainties in the
Galaxy morphological type and luminosity are not included.

It is also interesting to compute the expected number of SNe for the
``fiducial sample'' of galaxies (Tammann et al. \cite{tamm:94}) which
includes all galaxies from the RSA catalog with recession velocity
$\le 1200$ km~s$^{-1}$. By adding the individual contribution of each
galaxy we predict 28 SNe per decade. The fraction of SNe 
actually discovered in these galaxies depends on the intensity of SN
search programs and has been growing progressively from 2 SNe per
decade in the first half of the century, to 12 SNe in the period
1950-1975, to 20 SNe in most recent years. Given the uncertainties,
the last value is consistent with the claims that nowadays nearly all SNe
which explode in galaxies of the fiducial sample are discovered.

\subsection{Additional types of SNe}

Recently a few intrinsically faint SNIa (e.g. SN~1991bg) have been
discovered and the question raised if they could represent a separate
class or, instead, they are in the tail of a single luminosity
distribution for SNIa (the magnitude of SN~1991bg, $B=-16.34$, is
about $3\sigma$ fainter than the average for SN~Ia, Turatto et
al. \cite{91bg}). While the question is still open, it is clear that
there can be a severe bias against these faint SNe in magnitude
limited SN searches. 

By using the control time method we can give an
estimate of the rate of this possible class of SNIa assuming that:
$i)$ the average absolute magnitude for these SNe is that of
SN~1991bg; $ii)$ to this class belong also SNe 1986G, 1992K and
SN1991F (Turatto et al. \cite{91bg}). Note that all these SNe but
SN1991F are included in our SN sample.

For faint SN~Ia we obtain a rate of 0.05 SNu (averaged through all
galaxy types) implying that, although the observed faint Ia events are
only 5\% of the SNIa in our combined sample, actually they may  be
about 1/4 of all SNIa explosions (but not the majority of SN~Ia events 
as suggested by Schaefer \cite{schaefer}).

SN~II show a wide variety of behaviors. In the present calculation we
assumed that SN~II are a mixture of SN~IIL and IIP with the former
being about 0.5 mag brighter than the latter, and both showing wide
dispersions in absolute magnitudes.  With these assumptions we found that
the rates of the two subclasses are almost identical both peaking in
late spirals.

In this approach SN~1987A is considered as a normal type II plateau
reaching a peak luminosity at the lower end of the luminosity
distribution.  However it has been argued that SN~1987A could be the
representative of a separate class of faint SN~II most of which remain
undiscovered because of the low intrinsic luminosity (Schmitz \&
Gaskell \cite{gask}).  By assuming that this is the correct
interpretation, we can give an estimate of the rate of this subclass
of SNII, adopting as average parameters for this class those of
SN~1987A. The number of 87A-like SNII in our sample ranges from 1
(SN~1987A itself) to 3, if we accept SNe 1973R and 1982F as members of
this class (van den Bergh \& McClure \cite{vdb:89}).  These two SNe
were both as faint as SN~1987A, but their red colors were most likely
due to high extinction (Patat et al. \cite{patat2}) and they were
probably intrinsically ``normal'' SNII plateau.  Conservatively, we
estimate that the rate of 87A-like SNe in spirals ranges from 0.06 to
0.18 SNu, that is from 10\% to 30\% of the rate of ``normal'' SNII,
and are certainly not the preferred outcomes of core collapse.

Another possible separate subclass of SNII is that of type IIn
(Schlegel, \cite{schl}). SNe assigned to this class show an unusual profile
for the H$\alpha$ emission with a narrow peak sitting on a broad base
and no sign of P-Cygni absorption. It is believed that this feature
results from the interaction of the SN ejecta with a dense
circumstellar material (CSM). This could explain also the bright
absolute magnitude and very slow luminosity decline which has been
observed in some SN~IIn, e.g. SN~1988Z (Turatto et
al. \cite{88z}).  On the other side if the CSM is not so dense the
interaction may be weaker and the light curve could be very similar to
that of other SNII, as is the case of SN~1989C (Turatto et al. \cite{89c})

Two of the SNe included in our sample are classified IIn, namely SNe
1987B and 1987F. SN~1988Z, discovered independently both in the Asiago
and OCA searches, is not considered because the parent galaxy is not
listed in the RC3. Based on the two observed cases we estimates that
the rate of SNIIn in spirals is in the range 0.01-0.03 SNu (the
uncertainties result because of the variance in the photometric
behaviors).  This means that, although in recent years about 15-20\%
of the observed SN~II are classified IIn, actually they are only 2\%
to 5\% of SNII explosions.

\section{Conclusions}

The major improvement of the present work with respect to previous
determinations of the frequency of SNe is in the SN statistics. In
fact, our sample counts 110 SNe discovered by the five SN searches in
a sample of 7773 RC3 galaxies. Therefore, our SN statistics almost
doubles that of previous works (the rates in PII were based on 65 SNe)
and is even better than that of the so-called ``fiducial
sample'' which in  1990 counted 96 SNe (Tammann et
al. \cite{tamm:94}). Also we made an effort to update the
algorithm and to estimate the importance of the search biases, using
different approaches. 
Finally, it is important to note that by merging the databases of
SN searches with different characteristics we reduced the impact of
the assumptions on the input parameters, of the nuclear and
inclination bias corrections, and of other possibly hidden, search biases.
In this regard, especially important is the inclusion of the
Evans' visual search.  

The main results of the computation are the following:

\begin{enumerate}
\item the rate of SNIa in ellipticals, 0.13 SNu, is confirmed to be
lower than that in spirals;
\item the most prolific galaxies are late spirals where 2/3 of the SNe
are of type II (0.88 SNu);
\item SN Ib/c are relatively rare, being at most 40\% of all SN~I (0.16
SNu in late spirals);
\item if SNe 1991bg and 1987A are considered the prototypes of
separate classes of faint SNIa and SNII, respectively, the overall SN
rates should be raised only by $20-30$\%.  
\item The average frequency SN~IIn in spirals is $\le 0.03$ SNu. The fact
that they constitute about 20\% of all SNII presently discovered is
due to their high intrinsic brightness.
\end{enumerate}

\begin{acknowledgements}
We are grateful to S. van den Bergh and R.D. McClure for kindly
providing us the file with the Evans' observations. We also thank
A. Di Bartolomeo and G. Barbaro for useful discussion.  MH
acknowledges support provided for this work by the National Science
Foundation through grant number GF-1002-96 from the Association of
Universities for Research in Astronomy, Inc., under NSF Cooperative
Agreement No. AST-8947990 and from Fundaci\'{o}n Andes under project
C-12984. MH acknowledges also support by C\'{a}tedra Presidencial de
Ciencias (Chile) 1996-1997.
\end{acknowledgements}

\end{document}